\documentclass[aps,prl.reprint,twocolumn]{revtex4}
\usepackage{amsmath}
\usepackage{amsfonts}
\usepackage{amssymb}
\usepackage{ulem}
\usepackage{cancel}
\usepackage{graphicx}
\usepackage{wrapfig}
\usepackage{graphicx}
\usepackage{textcomp}
\usepackage[colorlinks=true, linkcolor=blue]{hyperref}
\newcommand{{\sign}}{\rm sign}

\newcommand{\mc}[1]{\mathcal{#1}}

\begin{document}	
	\title{%
		Quantum theory of wave mixing on a two-level system
	}

	\author{A. A. Elistratov$^{1}$}\email{andrei.a.elistratov@mail.ru}
	\author{S. V. Remizov$^{1,2,9}$}
	\author{W. V. Pogosov$,^{1,3,4}$}
	\author{A. Yu. Dmitriev$,^{5}$}
	\author{O. V. Astafiev$,^{5,6,7,8}$}

	\affiliation{$^1$Center for Fundamental and Applied Research, Dukhov Research Institute of Automatics (VNIIA),  Moscow 127055, Russia}
	\affiliation{$^2$V. A. Kotel'nikov Institute of Radio Engineering and Electronics, Russian Academy of Sciences, Moscow 125009, Russia}
 \affiliation{$^3$Center for Mesoscience and Nanotechnology, Moscow Institute of Physics and Technology (MIPT), Dolgoprudny, 141700, Russia}
\affiliation{$^4$Institute for Theoretical and Applied Electrodynamics,
Russian Academy of Sciences, Moscow, 125412, Russia} 
	\affiliation{$^5$Laboratory of Artificial Quantum Systems, Moscow Institute of Physics and Technology, 141700 Dolgoprudny, Russia}
	\affiliation{$^6$Physics Department, Royal Holloway, University of London, Egham, Surrey TW20 0EX, United Kingdom}
	\affiliation{$^7$National Physical Laboratory, Teddington, TW11 0LW, United Kingdom}
	\affiliation{$^8$Skolkovo Institute of Science and Technology, Nobel St. 3, 143026 Moscow, Russia}
        \affiliation{$^9$HSE University,  Moscow 109028, Russia}

\begin{abstract}
We apply the scattering matrix formalism to wave mixing on a quantum two-level system. We carry out the fermionization of the two-level system degrees of freedom using the Popov-Fedotov semions, calculate n-particle Green's function, and apply the Lehmann-Symanzik-Zimmermannn reduction procedure.
Using the developed approach, we provide a consistent quantum explanation of the appearance of coherent side peaks observed in an experiment on the scattering of bichromatic radiation on a two-level artificial atom \cite{dmitriev2019probing}. We show that the spectrum observed in the experiment is the result of bosonic stimulated scattering of photons from one mode of the bichromatic drive to another and vice versa.
\end{abstract}

\maketitle

\section{Introduction}

Rapid experimental progress in the field of waveguide QED systems motivated intensive theoretical efforts. The mutiphoton scattering matrix (S matrix), being one of the most important theoretical tools for studying the photon-photon interaction in waveguide QED systems, needs further development. 

Experimentally, there are several promising results arising out of the one-dimensional scattering problem \cite{roy2017colloquium}, where the scatterer is either a single superconducting artificial atom \cite{astafiev2010resonance,muller2007resonance} or multiple individually controlled or addressed atoms \cite{brehm2021waveguide,mirhosseini2019cavity}, strongly coupled to an open waveguide \cite{shen2007strongly,fan2010input}. This approach allows to study peculiar properties of atomic resonance fluorescence \cite{toyli2016resonance,campagne2014observing}, to observe electromagnetically induced transparency \cite{abdumalikov2010electromagnetically} and related phenomena \cite{sillanpaa2009autler}, to generate single photons on demand \cite{peng2016tuneable,zhou2020tunable}, to create and study photonic crystals \cite{brehm2022slowing}, cavities formed by strongly coupled atoms \cite{mirhosseini2019cavity}, and giant artificial atoms with distant coupling points \cite{kannan2020waveguide,kannan2023demand}. 

Among these, we focus on visualization of multi-photon scattering by careful collection of radiation coherently emitted by a single two-level atom \cite{dmitriev2017quantum,dmitriev2019probing,vasenin2022dynamics}. In a number of experiments, the case of either continuous \cite{dmitriev2019probing} or pulsed near-resonant \cite{dmitriev2017quantum,vasenin2022dynamics} bichromatic drive of a qubit consisting of mode $A$ with frequency $\omega_A$ and mode $B$ with frequency $\omega_B$ was examined, and it is shown that the coherent sidebands emitted by the scatterer characterize photon statistics of incoming pump fields.  The specifics of this regime is that the detuning of tones from each other and from exact resonance of qubit's transition is much smaller than the natural qubit's linewidth: two pumps interact equally strongly with the atom, however, they are perfectly distinguishable in field measurements as there is no unwanted absorption in resonance \cite{astafiev2010resonance}. Such a regime is not easily achievable in quantum optics with visible light irradiating atomic vapours \cite{zhu1990resonance,freedhoff1990resonance,ficek1993resonance}; in principle, it is achievable in quantum dots \cite{kryuchkyan2017resonance,he2019coherently,peiris2014bichromatic}, however as a platform for photonics, they do suffer from losses, matching problems and, in general, less controllable fabrication and lack of measurement control for individual quantum systems \cite{tomm2023photon}. Another advantage of supercondicting platform is that the wave dispersion is negligible \cite{shen2007strongly}, so the phase-matching conditions \cite{boyd2020nonlinear}, which are to be strictly satisfied to observe nonlinear wave mixing in optics \cite{fiore1998phase,kauranen2013freeing}, do not restrict the effectiveness of wave mixing on a single superconcting atom\cite{dmitriev2019probing}. 

Theoretically, the formalism of the S matrix as applied to the problems of quantum optics was consistently developed in several directions \cite{shen2007strongly,fan2010input, PhysRevB.79.205111,PhysRevA.82.063816,PhysRevA.84.063803,PhysRevA.91.043845,PhysRevA.92.053834,PhysRevA.95.043814,PhysRevA.95.053845,PhysRevLett.106.053601,PhysRevA.87.063819,PhysRevA.95.053845,PhysRevA.82.053836,PhysRevA.83.063828,PhysRevLett.106.113601,PhysRevA.87.043809}. 
The wave-function approach for photon S matrix calculation arose \cite{shen2007strongly,PhysRevLett.106.053601,PhysRevA.87.063819,PhysRevA.82.053836,PhysRevA.83.063828,PhysRevLett.106.113601,PhysRevA.87.043809}. In particular, within the framework of this approach, using the Bethe ansatz \cite{shen2007strongly}, the completeness of the photon S matrix was shown. An alternative approach in the spirit of quantum field theory is that the n-particle Green's function is calculated, and then the Lehmann-Symanzik-Zimmermannn (LSZ) reduction procedure is applied \cite{PhysRevB.79.205111}. This approach, originally created for the scattering of a small number of photons, was generalized to the scattering of continuous-mode wave packets, in particular, coherent-state wave packets \cite{PhysRevA.82.063816}.
In addition, the relationships between the S matrix formalism and the method of the density matrix master equation including in-out theory, which is more commonly used in quantum optics, were investigated \cite{Caneva_2015,fan2010input,PhysRevA.91.043845,PhysRevA.92.053834}. 

In this paper, we apply the S matrix formalism to the problem of wave mixing on a quantum two-level system (TLS).
Three- and four-wave mixing is a long-known optical phenomenon in nonlinear media, described purely classically in terms of electrodynamics. The development of experimental possibilities made available the study of light scattering on a single TLS, in particular, microwave radiation on an artificial atom. Such a process is usually described with the help of the semiclassical approach, in which the degrees of freedom of the TLS are considered quantum, while the pump fields are considered classical \cite{Mandel}. This approach, as applied to the problem under consideration, made it possible to describe the appearance of a side peaks and the dependence of their amplitudes on the pump intensity \cite{dmitriev2019probing,Pogosov}. At the same time, further improvement of experimental equipment, in particular, the appearance of single-photon sources and receivers, made a purely quantum description of photon scattering processes on a single TLS relevant. One of the variants of the quantum description is the formalism of the scattering matrix developed in the framework of quantum field theory \cite{shen2007strongly,PhysRevB.79.205111,PhysRevA.82.063816,PhysRevA.84.063803,PhysRevA.91.043845,PhysRevA.92.053834,PhysRevA.95.043814,PhysRevA.95.053845}. In this paper, we use this approach to construct the quantum theory of  wave mixing on a quantum two-level  system. 

In more detail, in order to use path integral formalism, we carry out the fermionization of the TLS degrees of freedom using the Popov-Fedotov semions, whose Green's functions are written in the Matsubara representation. 
We show how to pass from the temperature Green's function used in many-body physics to the $n$-particle zero-temperature Green's function used in $S$-matrix theory. We derive an explicit expression for the diagrams that give the leading contribution to the amplitude of a side  peak. Comparing combinatorial prefactors of the obtained diagrams, we see that the most probable scattering channel is bosonic stimulated scattering from mode $A$ to mode $B$ and vice versa.  Thus, we show that the spectrum observed in the experiment is the result of bosonic stimulated scattering of photons from mode $A$ to mode $B$ for the side peaks on the right, and from mode $B$ to mode $A$ for the side peaks on the left of two initial peaks with frequencies $\omega_A$ and $\omega_B$. 

The paper is organized as follows. In the next section, we briefly discuss the experiment that prompted us to undertake the construction of the presented theory. Section~III presents a semiclassical approach to the interpretation of the experiment, in which the TLS degrees of freedom are considered quantum, and the pumps are considered classical.  In Section~IV we derive the effective action of the system and use it in Section~V to obtain the scattering matrix. In Section ~VI we return to the experiment, calculate the amplitudes of the side peaks and show how they turn into semiclassical expressions obtained in Section~III. We conclude in Section~VII.    

\section{Experiment}

\begin{figure}[htp] 
	\includegraphics[width=0.95\columnwidth]{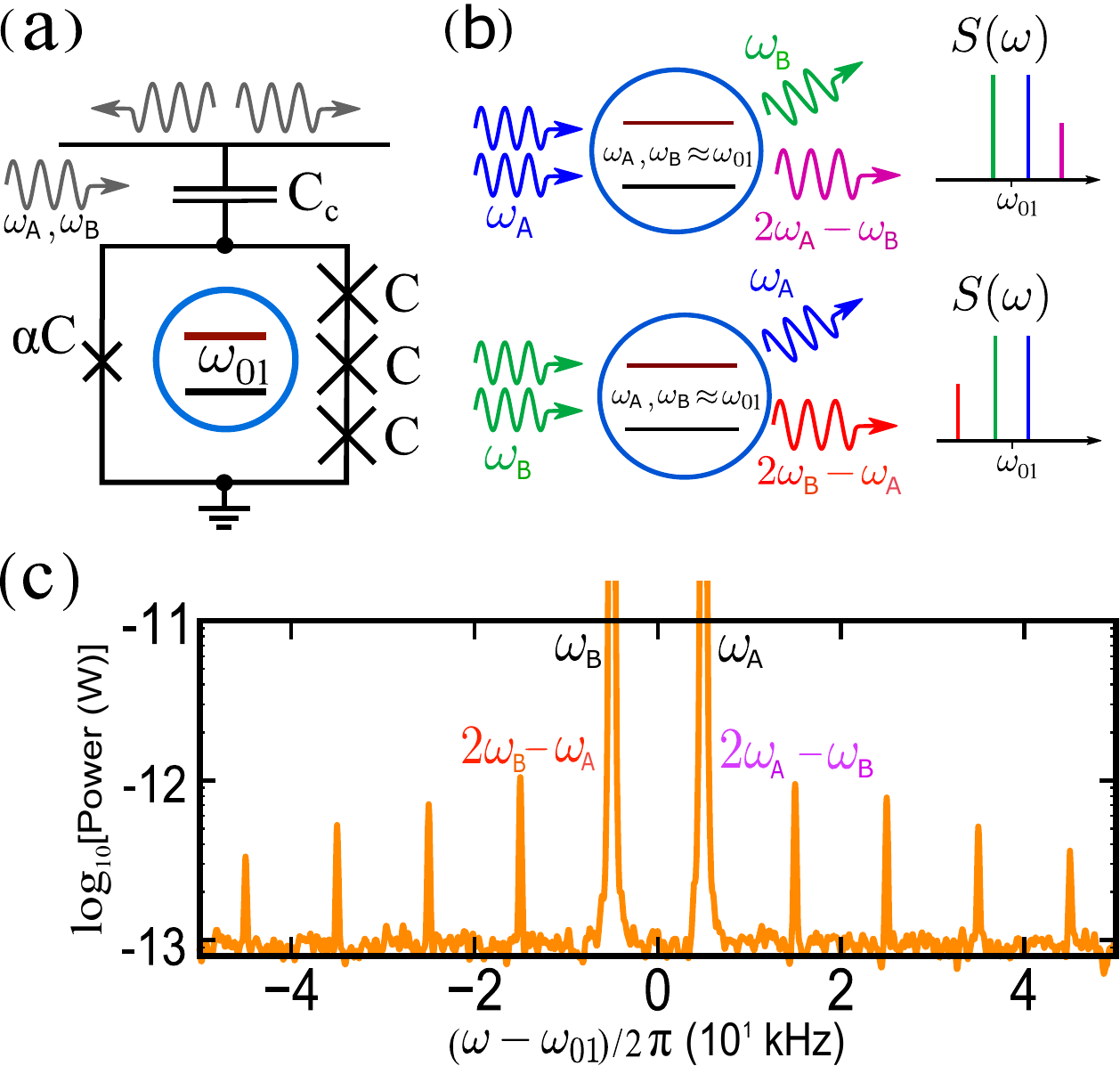}
	\caption{\small (a) Schematics of the device (see details in \cite{dmitriev2019probing}). (b) The wave mixing of two tones on a single two-level system. With the tho-photon absorption and one-photon emission at frequencies $\omega_{A,B}$ and $\omega_{B,A}$ one more photon is generated at frequency $2\omega_{A,B}-\omega_{B,A}$. (c) An example of a typical spectrum.  
		\label{Fig_experiment}
	}
\end{figure}	 
The physical system of study is sketched in Fig.~\ref{Fig_experiment}a. A two-level artificial atom with the transition energy $\hbar\omega_{01}$ is coupled with a a microwave line, the radiative relaxation rate is $\Gamma$. We neglect the non-radiative relaxation of the atom. The artificial atom is a 4-junction flux qubit (see details in \cite{dmitriev2019probing}).  Two monochromatic waves with a linear dispersion law $\omega=vk$ and frequencies $\omega_A=\omega_{01}+\delta\omega$ and  $\omega_B=\omega_{01}-\delta\omega$, where $\delta\omega\ll \Gamma$, propagate along the line and scatter on the atom. Concerning the scattered radiation, it is known that, in addition to two initial peaks at frequencies $\omega_A$ and $\omega_B$, a number of narrow side  peaks appear (see Fig.~\ref{Fig_experiment}c). Namely, one could observe peaks which frequencies to the right of  $\omega_{01}$ are equal to $\omega^{(2p+1)}=(p+1)\omega_A-p\omega_B=\omega_{01}+(2p+1)\delta\omega$, and to the left of $\omega_{01}$ are $\omega^{(2p+1)}=(p+1)\omega_B-p\omega_A=\omega_{01}-(2p+1)\delta\omega$, where $p$ is a positive integer. Each of the side  peaks can be interpreted as a manifestation of the elastic $(p+1)$-photon scattering of one of the modes, as illustrated in Fig.~\ref{Fig_experiment}b. As a result, $p$ photons of the counterpart mode are born. In addition, one photon is born with a frequency following from the energy conservation law. These photons form the $(2p+1)$th side  peak. 

The experimental observation of side peaks \cite{dmitriev2019probing} is possible under several conditions. The side-coupled atom emits all the radiation symmetrically, forwards and backwards, except the elastically scattered Rayleigh component at the frequency of drive, which interferes with the drive and therefore could exhibit either good transmission or good reflection, depending on the driving amplitude. The scattered field, either transmitted or reflected, should be effectively amplified and collected either by a linear field detector or by spectral analyzer. In both cases, the effective resolution bandwidth is to be much smaller than $\delta\omega$, and this is easily achieved with typical radio-frequency detectors. The pure dephasing rate of the atom is required to be either negligible or, at least, comparable to its radiational decay rate. The aforementioned side components are a manifestation of atom's saturation under strong drive, therefore, effective Rabi frequency of each of coherent drives is required to be comparable to the $\Gamma$.

\section{Semiclassical treatment}
\label{ST}
The semiclassical approach was considered in detail in \cite{dmitriev2019probing, Pogosov}, so here we just briefly recall its main results.

The Hamiltonian of a TLS subjected to two classical pumps $A$ and $B$ can be written in the form 
\begin{equation}
\mc{H}_{\rm cl}=-\frac{\varepsilon}{2}\sigma_z-\sigma_x\hbar\Omega_A\cos\omega_At-\sigma_x\hbar\Omega_B\cos\omega_Bt
\end{equation}
The quasi-stationary solution of the equations of motion has the form 
\begin{multline}
\langle\sigma^-\rangle=-\Gamma\frac{\sin\theta}{4\Omega_A\Omega_B}\frac{\Omega_Ae^{i\delta\omega t}+\Omega_Be^{-i\delta\omega t}}{1+\sin\theta\cos2\delta\omega t}\\
=-\Gamma\frac{\Omega_Ae^{i\delta\omega t}+\Omega_Be^{-i\delta\omega t}}{4\Omega_A\Omega_B}\tan\theta\sum_{p=-\infty}^{\infty}y^{|p|}e^{i2p\delta\omega t},
\end{multline}
where
\begin{equation}
\theta=\arcsin\frac{\Omega_A\Omega_B}{\Gamma^2/4+(\Omega_A^2+\Omega_B^2)/2},\quad y=-\tan\frac{\theta}{2}. 
\end{equation}

The amplitudes of the side peaks can be found using the relation $\Omega=-i\Gamma\langle\sigma^-\rangle$. As a result, we get the amplitude of the $(2p+1)$th side peak as
\begin{equation}
\Omega^{(2p+1)}=(-1)^p\Gamma^2\frac{\tan\theta\tan^p\frac{\theta}{2}}{4\Omega_A\Omega_B}\left\{\begin{aligned}\Omega_B\tan\frac{\theta}{2}-\Omega_A&\quad\text{(r),}\\\Omega_A\tan\frac{\theta}{2}-\Omega_B&\quad\text{(l).}\end{aligned}\right.
\end{equation}
Here and below (r) denotes side peaks on the right, (l) denotes side peaks on the left of two initial peaks with frequencies $\omega_A$ and $\omega_B$.

In the limit of small amplitudes $\Omega_A$ and $\Omega_B$, this relation reduces to   
\begin{equation}
\Omega^{(2p+1)}=(-1)^{p+1}\frac{2^p}{\Gamma^{2p}}\left\{\begin{aligned}\Omega_A^{p+1}\:\Omega_B^p&\quad\text{(r),}\\\Omega_A^p\:\Omega_B^{p+1}&\quad\text{(l).}\end{aligned}\right.
\end{equation}
Introducing the number of photons in the mode $N=\sqrt{2}|\Omega|/\Gamma$, we obtain the relation 
\begin{equation}\label{N1}
N^{(2p+1)}=\left\{\begin{aligned}N_A^{p+1}\:N_B^p&\quad\text{(r),}\\N_A^p\:N_B^{p+1}&\quad\text{(l).}\end{aligned}\right.
\end{equation}
The interpretation of side  peaks proposed in the previous section as manifestation of $(p+1)$-photon elastic scattering processes in no way clarifies the appearance of relation (\ref{N1}). Moreover, such an interpretation, provided that the pumps  $A$ and $B$ are considered classically as $c$-numbers, is hardly appropriate. In the following sections, we construct a fully quantum theory of wave mixing on a two-level system and obtain relation  (\ref{N1}) as the classical limit achievable for large $N_A$ and $N_B$.

\section{Action}
In this section, we get away from the experiment and consider the problem from a more general point of view.

\label{Action}
\subsection{Hamiltonian}
\begin{figure}[htp] 
	\includegraphics[width=0.5\columnwidth]{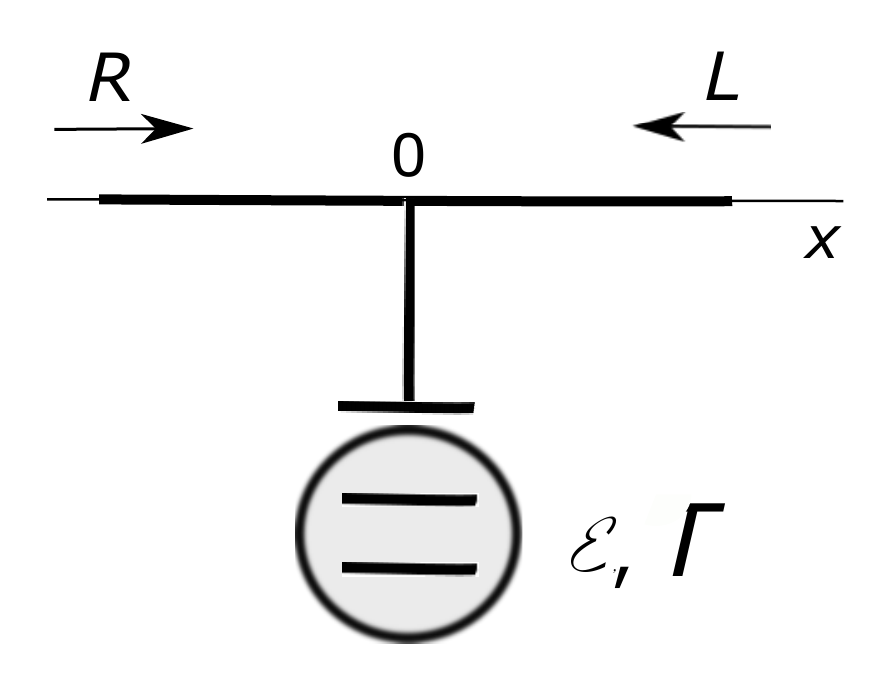}
	\caption{\small The TLS coupled with the waveguide. 
		\label{Fig_Q}
	}
\end{figure}

The system we study consists of a quantum two-level system with the transition energy $\varepsilon$ coupled with a 1D waveguide, the radiative relaxation rate is $\Gamma$ (see Fig.~\ref{Fig_Q}). We neglect non-radiative relaxation of the system. Photons propagate along the waveguide in both directions and scatter on the TLS located at the point $x=0$. The system is described by the Hamiltonian

\begin{multline}
\mc{H}=\frac{\varepsilon}{2}\sigma_z\\
+\int dx\left\{\hbar\left[-i v c_R^+(x)\frac{\partial}{\partial x}c_R(x)+i v c_L^+(x)\frac{\partial}{\partial x}c_L(x)\right]\right.\\
+\left.V\delta(x)[c_R^+(x)\sigma_-+c_R(x)\sigma_++c_L^+(x)\sigma_-+c_L(x)\sigma_+]\right\},
\end{multline}
where operators $\sigma_-$ and $\sigma_+$  provide transition between the ground and excited levels of the TLS, $c_R^+(x)$ and $c_L^+(x)$ are the creation operators for, respectively, a right-going and left-going photon at position $x$, $v$ is the speed of photons in the waveguide. 

In order to avoid consideration of the transformations of $R$-photons into $L$-photons and vice versa, we pass to $e$-photons with even parity in the momentum space, i.e., $c_{-k,e}=c_{k,e}$, and $o$-photons with odd parity i.e., $c_{-k,o}=-c_{k,o}$, making a standard rotation of the basis \cite{shen2007strongly}
\begin{align}\label{RLeo}
c_e^+(x)=&\frac{1}{\sqrt{2}}[c_R^+(x)+c_L^+(-x)],\nonumber\\
c_o^+(x)=&\frac{1}{\sqrt{2}}[c_R^+(x)-c_L^+(-x)].
\end{align}
In the new variables, the Hamiltonian takes the form $\mc{H}=\mc{H}_e+\mc{H}_o$, where $e$ is a chiral (non-scattering backwards) mode entering the Hamiltonian
\begin{multline}\label{H_e}
\mc{H}_e=\frac{\varepsilon}{2}\sigma_z+\int dx\left\{-i v \hbar c_e^+(x)\frac{\partial}{\partial x}c_e(x)\right.\\
\left.+\tilde V\delta(x)[c_e^+(x)\sigma_-+c_e(x)\sigma_+]\right\},
\end{multline}
and $o$ is the mode that does not interact with the TLS and has the Hamiltonian
\begin{equation}
\mc{H}_o=\int dx(-i) v\hbar c_o^+(x)\frac{\partial}{\partial x}c_o(x).
\end{equation}
Further, we can deal only with the scattering of $e$-photons, remembering about $o$-photons when returning to the original basis.

\subsection{Popov-Fedotov semions}
The application of path integration at the next step requires fermionization of the TLS degrees of freedom.
We rewrite the variables of the TLS in terms of fermions $a$ and $b$ as \cite{Abrikosov}
\begin{equation}\label{APF}
\hat\sigma_z=\hat a^+\hat a-\hat b^+\hat b,\quad \hat\sigma_+=\hat a^+\hat b ,\quad \hat\sigma_-=\hat b^+\hat a.
\end{equation}
The bare Matsubara Green's functions of fermions $a$ and $b$ are given by the relations
\begin{align}\label{GaGbM}
G_a^{(0)}(i\omega)=&\frac{1}{i\omega-(\varepsilon/2-\mu)/\hbar},\nonumber\\ G_b^{(0)}(i\omega)=&\frac{1}{i\omega-(-\varepsilon/2-\mu)/\hbar},
\end{align}
where $\omega=\omega_m=2\pi(m+1/2)/\hbar\beta$ are the Matsubara fermionic frequencies, and also introduced the imaginary chemical potential $\mu=-i\pi/2\beta$, where $\beta=1/k_BT$, $T$ is the temperature \cite{Popov}. One may include the chemical potential $\mu$ in the frequencies $\omega_m$ and obtain frequencies $\tilde\omega_m=2\pi(m+1/4)/\hbar\beta$, which are intermediate betweeen bosonic and fermionic ones. For this reason, such fermions are called semions. The imaginary chemical potential introduced by Popov and Fedotov makes it possible to eliminate the TLS non-physical states $|00\rangle$ and $|11\rangle$ in an explicit form (see \cite{Elistratov} for detailes), and not to sweep the corresponding constraint under the carpet of the measure of path integration. In addition, the fact that semion Green's functions are written in the Matsubara formalism allows us to consider the problem of photon scattering by a TLS coupled with a reservoir at a finite temperature. 

\subsection{Effective action}
So, having made all necessary substitutions, we can proceed to calculate the action of the system. 
The action corresponding to Hamiltonian (\ref{H_e}) can be written in the Matsubara technique as follows
\begin{equation}\label{A}
\mc{A}=\mc{A}_{\rm TLS}+\mc{A}_{\rm int}+\mc{A}_{\rm c},
\end{equation}
where $\mc{A}_{\rm TLS}$ describes the TLS
\begin{multline}
\mc{A}_{\rm TLS}=\int_0^{\hbar\beta} d\tau \left[\bar a(\tau)(\hbar\partial_\tau+\varepsilon/2-\mu)a(\tau)\right.\\
\left.+\bar b(\tau)(\hbar\partial_\tau-\varepsilon/2-\mu)b(\tau)\right],
\end{multline}
$\mc{A}_{\rm int}$ describes the interaction between the TLS and the resonator
\begin{multline}
\mc{A}_{\rm int}=\int_0^{\hbar\beta} d\tau\int dx \:\tilde V\delta(x)\left[ \bar c_e(x,\tau)\bar b(\tau)a(\tau)\right.\\
\left.+c_e(x,\tau)\bar a(\tau)b(\tau)\right],
\end{multline}
and $\mc{A}_{\rm c}$ describes the resonator
\begin{equation}
\mc{A}_{\rm c}=\hbar\int_0^{\hbar\beta} d\tau\int dx\: \bar c_e(x,\tau)(\partial_\tau-iv\partial_x)c_e(x,\tau).
\end{equation}

Let us pass to the momentum representation performing the Fourier transformation $c_e(x,\tau)=\int e^{ikx}c_e(k,\tau)dk/2\pi$. We obtain
\begin{multline}
\mc{A}_{\rm int}=\int_0^{\hbar\beta} d\tau\int\frac{dk}{2\pi}\:\tilde V\left[ \bar c_e(k,\tau)\bar b(\tau)a(\tau)\right.\\
\left.+c_e(k,\tau)\bar a(\tau)b(\tau)\right],
\end{multline}
\begin{equation}\label{Ac}
\mc{A}_{\rm c}=\hbar\int_0^{\hbar\beta} d\tau\int \frac{dk}{2\pi}\;\: \bar c_e(k,\tau)(\partial_\tau+vk)c_e(k,\tau).
\end{equation}

The Green's function of the photons $G(k,\tau_1-\tau_2)=-\langle \mc{T}_\tau c_e(k,\tau_1)c_e^+(k,\tau_2)\rangle$, where $\mc{T}_\tau$ stands for time-ordering, can be found  by varying the generating functional 
\begin{multline}\label{Gkt}
G(k,\tau_1-\tau_2)\\
=\left.-\frac{1}{2\pi}\frac{\delta^2}{\delta \bar\eta(k,\tau_1)\:\delta \eta(k,\tau_2)}\frac{\int \mc{D}\Theta \:e^{-\mc{A}[\eta,\bar\eta]/\hbar}}{\int \mc{D}\Theta \:e^{-\mc{A}/\hbar}}\right|_{\eta, \bar\eta=0},
\end{multline}
in which
\begin{multline}\label{Anunu}
\mc{A}[\eta,\bar\eta]=\mc{A}-\hbar\int_0^{\hbar\beta} d\tau\int dk\: [c_e(k,\tau)\bar\eta(k,\tau)\\
+\eta(k,\tau)\bar c_e(k,\tau)],
\end{multline}
$\int \mc{D}\Theta$ denotes taking the path integral over all fields involved in the action.

Let us now do the shift of variables in the action 
\begin{equation}
c_e(k,\tau)\to c_e(k,\tau)-\int_0^{\hbar\beta} d\tau'\: G^{(0)}(k,\tau-\tau')\:\eta(k,\tau'),
\end{equation}
where $G^{(0)}(k,\tau-\tau')$ is the inverse of
$G^{(0)-1}(k,\tau-\tau')=-(\partial_\tau+vk)\delta(\tau-\tau')$, so
\begin{multline}
\int_0^{\hbar\beta} G^{(0)-1}(k,\tau-\tau_1)G^{(0)}(k,\tau_1-\tau')d\tau_1\\
=-(\partial_\tau+vk)G^{(0)}(k,\tau-\tau')=\delta(\tau-\tau'),
\end{multline}
or in matrix form  $G^{(0)-1}G^{(0)}=I$, $I$ is the identity matrix.

We pass to the frequency representation using the Fourier transformation $c_e(k,\tau)=\sum_\omega e^{-i\omega \tau}c_e(k,i\omega)/\hbar\beta$. Formula (\ref{Ac}) takes the form
\begin{equation}
\label{Ac1}
\frac{\mc{A}_{\rm c}}{\hbar}=\frac{1}{\hbar\beta}\sum_\omega\int \frac{dk}{2\pi}\;\: \bar c_e(k,i\omega)(-i\omega+vk)c_e(k,i\omega),
\end{equation}
while formula (\ref{Anunu}) goes into 
\begin{multline}
\label{Aetaeta}
\frac{\mc{A}[\eta,\bar\eta]}{\hbar}=\frac{\mc{A}}{\hbar}-\sum_\omega d\tau\int dk\: \left[c_e(k,i\omega)\frac{\bar\eta(k,i\omega)}{\hbar\beta}\right.\\
\left.+\frac{\eta(k,i\omega)}{\hbar\beta}\bar c_e(k,i\omega)\right].
\end{multline}
It follows from the relations (\ref{Ac1}) and (\ref{Aetaeta}) that the Fourier component of the function (\ref{Gkt}), which is determined by the relation $G(k,i\omega)=\int_0^{\hbar\beta}e^{i\omega\tau}G(k,\tau)d\tau$, can be calculated using the expression
\begin{multline}\label{Gkw}
G(k,i\omega)=-\langle c_e(k,i\omega)c_e^+(k,i\omega)\rangle\\
=\left.-\frac{\hbar\beta}{2\pi}\:\frac{\delta^2}{\delta \bar\eta(k,i\omega)\:\delta \eta(k,i\omega)}\frac{\int \mc{D}\Theta \:e^{-\mc{A}[\eta,\bar\eta]/\hbar}}{\int \mc{D}\Theta \:e^{-\mc{A}/\hbar}}\right|_{\eta, \bar\eta=0}.
\end{multline}
As a result, after the shift and transition to the frequency representation, the action takes the form
\begin{widetext}
\begin{multline}
\mc{A}[\eta,\bar\eta]=-\sum_\omega\left(\frac{1}{\beta}\int \frac{ dk}{2\pi}\: \bar c_e(k,i\omega)(i\omega-vk)c_e(k,i\omega)-\frac{2\pi}{\beta}\int  dk\:  \bar \eta(k,i\omega)\: G^{(0)}(k,i\omega)\:\eta(k,i\omega)+\right.\\
\left.+\frac{1}{\hbar\beta}\left[\bar a(i\omega)(i\hbar\omega-\varepsilon/2+\mu)a(i\omega)+\bar b(i\omega)(i\hbar\omega+\varepsilon/2+\mu)b(i\omega)\right]\right)+\\
+\sum_{\omega_1, \omega_2}\left(\bar a(i\omega_1)\:\frac{\tilde V}{(\hbar\beta)^2} \int dk  \left[G^{(0)}(k,i\omega_1-i\omega_2)\:\eta(k,i\omega_1-i\omega_2)-\frac{c_e(k,i\omega_1-i\omega_2)}{2\pi}\right]\:b(i\omega_2)+\right.\\
\left.+\bar b(i\omega_1)\:\frac{\tilde V}{(\hbar\beta)^2} \int dk  \left[G^{(0)}(k,i\omega_1-i\omega_2)\:\bar\eta(k,i\omega_1-i\omega_2)- \frac{\bar c_e(k,i\omega_1-i\omega_2)}{2\pi}\right]\:a(i\omega_2)\right).
\end{multline}
\end{widetext}
Here
\begin{equation}
G^{(0)}(k,i\omega)=\frac{1}{i\omega-\omega_k},\quad \omega_k=vk.
\end{equation}
Integration over the semion fields $a/\sqrt{\hbar\beta}$ and $b/\sqrt{\hbar\beta}$ leads to the effective action
\begin{multline}\label{Aeff}
\frac{\mc{\tilde A}_{\rm eff}[\eta,\bar\eta]}{\hbar}=-\frac{1}{\hbar\beta}\sum_\omega \left(\int   \frac{ dk}{2\pi}\:  \bar c_e(k,i\omega)(i\omega-vk)c_e(k,i\omega)\right.\\
\left.-2\pi\int  dk\: \bar \eta(k,i\omega)\: G^{(0)}(k,i\omega)\:\eta(k,i\omega)\right)+\\
+{\rm Tr}\ln {\bf \tilde M},
\end{multline}
where
\begin{multline}\label{M}
{\bf \tilde M}=\begin{bmatrix} \left(i\omega_1-\frac{\varepsilon/2-\mu}{\hbar}\right)\delta_{0,i\omega_{12}} &  \tilde\Phi(\omega_1,\omega_2) \\ \bar\tilde\Phi(\omega_1,\omega_2) & \left(i\omega_1-\frac{-\varepsilon/2-\mu}{\hbar}\right)\delta_{0,i\omega_{12}} \end{bmatrix},\\
\tilde\Phi(\omega_1,\omega_2)=g\frac{2\pi}{\hbar\beta}\int \frac{dk}{2\pi} G^{(0)}(k,\omega_{12})\:\eta(k,\omega_{12})\\
-g\frac{1}{\hbar\beta}\int \frac{dk}{2\pi}c_e(k,i\omega_{12}).
\end{multline}
Here $\omega_{12}=\omega_1-\omega_2$ and $g=\tilde V/\hbar$.

Further, one can represent  $\ln {\bf \tilde M}$ as $\ln(1+{\bf G}^{(0)}{\bf\Phi})$ in the standard way, where
\begin{multline}
{\bf G}^{(0)}={\bf G}^{(0)}(\omega_1,\omega_2)=\,\delta(\omega_{12})\begin{bmatrix} G_a^{(0)}(\omega_1) & 0 \\ 0 & G_b^{(0)}(\omega_1) \end{bmatrix},\\
{\bf\tilde \Phi}={\bf \Sigma}(\omega_1,\omega_2)=\begin{bmatrix} 0 & \tilde\Phi(\omega_1,\omega_2) \\ \tilde\bar\Phi(\omega_2,\omega_1) & 0 \end{bmatrix},\\
\end{multline}
and use the Taylor series expansion of the logarithm $\ln(1+{\bf G}^{(0)}{\bf\Phi})= \sum_{j=1}^\infty \frac{(-1)^j}{j}({\bf G}^{(0)}{\bf\Phi})^j$. In addition, we use the relation
\begin{equation}
 {\rm Tr}K^k=\int dz_1\dots dz_k K(z_1,z_2) K(z_2,z_3)\dots K(z_k,z_1).
\end{equation}

The resulting effective action (\ref{Aeff}) allows us to consider many problems besides the scattering problem. So, in the next subsection, we will take a step aside and consider the diagrammatic approach to dressing of the TLS Green's function with an interaction with a waveguide.

\subsection{TLS Green's function}
Let us now see what results expression (\ref{Aeff}) leads to in different orders of perturbation theory. The first term of the sum $(j=1)$ is equal to zero. The second term of the sum $(j=2)$ is equal to $\frac{1}{2}{\rm Tr}(G_a^{(0)}\Phi G_b^{(0)}\bar\Phi+G_b^{(0)}\bar\Phi G_a^{(0)}\Phi)$, which allows us to write the first term on the right-hand side of (\ref{Aeff}) in a refined form  
\begin{multline}
-\int \frac{ dk}{2\pi}\frac{ dk'}{2\pi}\:\frac{1}{\hbar\beta}\sum_\omega \bar c_e(k,i\omega)\left(G^{(0)-1}(k,i\omega)2\pi\delta(k-k')\right.\\
\left.-g^2\frac{1}{\hbar\beta}\sum_{\omega_1} G_a^{(0)}(i\omega_1) G_b^{(0)}(i\omega_1-i\omega)\right)c_e(k',i\omega)
\end{multline}
and make an equivalent correction to the second term of the right-hand side 
\begin{multline}\label{G1}
\int \frac{ dk}{2\pi}\frac{ dk'}{2\pi}\:\frac{1}{\hbar\beta}\sum_\omega\bar \eta(k,i\omega)\: 
\Bigg(G^{(0)}(k,i\omega)(2\pi)^3\delta(k-k')+\\
+(2\pi)^2 G^{(0)}(k,i\omega)\\
\left.\times\frac{g^2}{\hbar\beta}\sum_{\omega_1}G_a^{(0)}(i\omega_1) G_b^{(0)}(i\omega_1-i\omega)G^{(0)}(k',i\omega)\right)\:\eta(k',i\omega).
\end{multline}

As we can see, the construction  $1/(\hbar\beta)\sum_{\omega_1} G_a^{(0)}(i\omega_1) G_b^{(0)}(i\omega_1-i\omega)$, which we denote as $G_q^{(0)}(i\omega)$, arises. It can be easily summed over the frequency $\omega_1$ by the partial fraction decomposition and formula 
\begin{equation}
\lim_{\eta\to 0}\:\frac{1}{\hbar\beta}\sum_{m=-\infty}^{m=+\infty}\frac{e^{-i\omega_m\eta}}{i\omega_m-(\varepsilon-\mu)/\hbar}=\frac{1}{e^{\beta(\varepsilon-\mu)}+1},
\end{equation} 
in which the summation is over the fermionic Matsubara frequencies $\omega_m=2\pi(m+1/2)/\hbar\beta$. We have
\begin{multline}
\label{Gq0}
G_q^{(0)}(i\omega)=\frac{1}{\hbar\beta}\sum_{\omega_1} G_a^{(0)}(i\omega_1) G_b^{(0)}(i\omega_1-i\omega)\\
=\frac{1}{i\omega-\varepsilon/\hbar}\left(\frac{1}{e^{\beta(-\varepsilon/2-\mu)}+1}-\frac{1}{e^{\beta(\varepsilon/2-\mu)}+1} \right)\\
=\frac{{\rm tanh}(\beta\varepsilon/2)}{i\omega-\varepsilon/\hbar}.
\end{multline}
Further, we restrict ourselves to the limit of low temperatures $\beta\to\infty$, for which ${\rm tanh}(\beta\varepsilon/2)=1$. We can call $G_q^{(0)}(i\omega)$ the bare Green's function of the TLS.
\begin{figure}[htp] 
	\includegraphics[width=0.6\columnwidth]{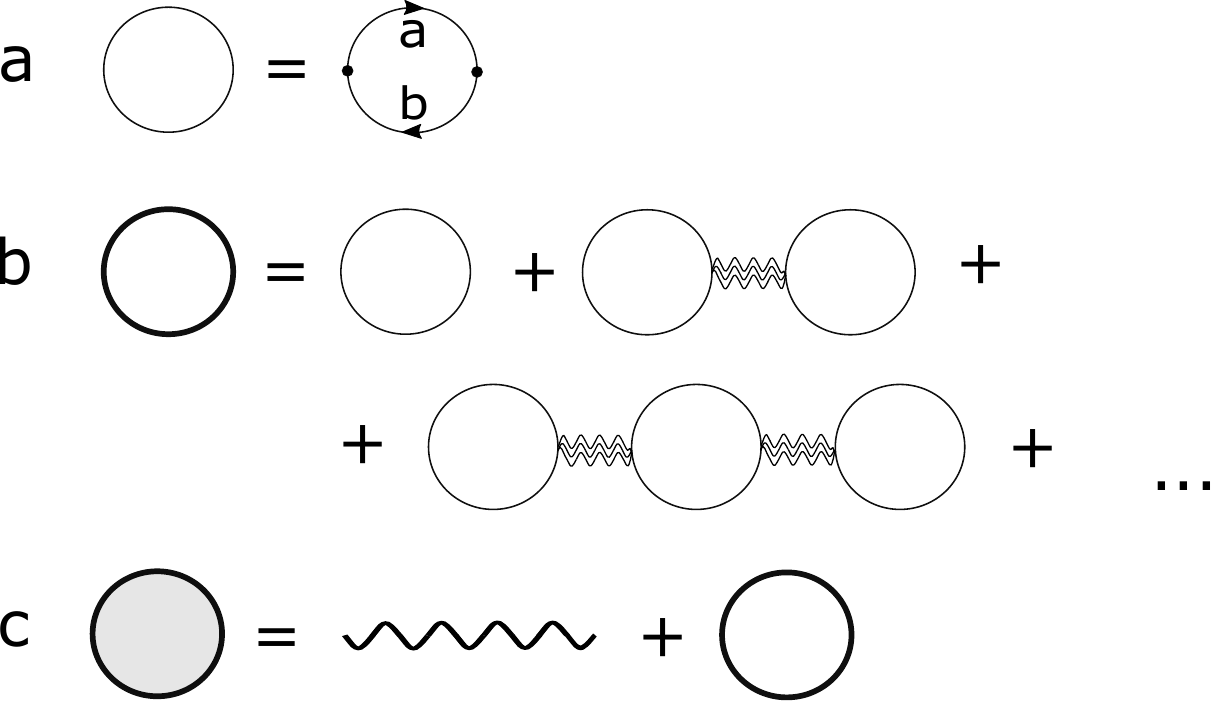}
	\caption{\small  Single-photon scattering. (a) The lines $a$ and $b$ in the right-hand side denote the semion Green's functions $G_a^{(0)}$ and $G_b^{(0)}$ correspondently, the open circle in the left-hand side denotes the bare TLS Green's function $G_q^{(0)}$ given by (\ref{Gq0}). (b)  The bold circle in the left-hand side denotes the function $G_q$ given by (\ref{Gq}) which is a result of dressing 
 the bare function $G_q^{(0)}$ by the interaction with the waveguide depicted by triple wavy lines in the right-hand side. (c) The gray circle denotes the full single-photon S matrix (see Section \ref{SPS}).  
		\label{Fig_OPD}
	}
\end{figure}	 

Let us discuss the corresponding diagrams. If the semion Green's functions $G_a^{(0)}$ and $G_b^{(0)}$  are represented by lines, then $G_q^{(0)}$ is represented by a circle, as shown in Fig. ~\ref{Fig_OPD}a. The forth order ($j=4$) adds to expression  (\ref{G1}) a term containing the element 
\begin{equation}
G_q^{(0)}(i\omega)\:g^2\int\frac{ dk_1}{2\pi}G^{(0)}(k_1,i\omega)\:G_q^{(0)}(i\omega),
\end{equation} 
which can be depicted as two circles, connected by a triple wavy line depicting the element $g^2\int G^{(0)}(k_1,i\omega) dk_1/2\pi$. Each next order of perturbation theory will give a similar contribution, which is an increasing in length chain of circles connected by triple interaction lines (Fig.~\ref{Fig_OPD}b). This sequence of diagrams can be summed in the standard way, which gives the TLS Green's function dressed by interaction with the waveguide
\begin{equation}\label{Gq}
G_q(i\omega)=\frac{1}{G_q^{(0)-1}-\Sigma(i\omega)}=\frac{1}{i\omega-\varepsilon/\hbar-\Sigma(i\omega)},
\end{equation} 
where
\begin{equation}\label{Sigma}
\Sigma(i\omega)=g^2\int\frac{ dk}{2\pi}G^{(0)}(k,i\omega)=g^2\int\frac{ dk}{2\pi}\frac{1}{i\omega-\omega_k}.
\end{equation} 
Making an analytical continuation $i\omega\to \omega+i0 $ we can rewrite this integral as
\begin{multline}
g^2 P\int \frac{ dk}{2\pi}\frac{1}{\omega-\omega_k}-i\pi g^2\int\frac{ dk}{2\pi}\delta(\omega-\omega_k)\\
=g^2 P\int \frac{ dk}{2\pi}\frac{1}{\omega-\omega_k}-\frac{g^2}{2}\sum_j\frac{1}{|\partial\omega_k/\partial k|_j}.
\end{multline} 
Here the subscript $j$ enumerates the zeros of the delta function $\delta(\omega-\omega_k)$ argument. We can consider different types of thermal reservoirs characterized by different density of states $|\partial\omega_k/\partial k|$, for example, ohmic \cite{Elistratov}.   
The simplest and most commonly used approach is to consider an open waveguide as a reservoir with a linear dispersion law $\omega_k=vk$ and a continuum distribution of states. In this case, we can represent $\Sigma(i\omega)$ as  	$\delta\varepsilon/\hbar-i\Gamma/2$, where   $\delta\varepsilon$ is considered as already included in $\varepsilon$, and $\Gamma=g^2/v$. Thus, we assume that taking into account the interaction of the TLS with a reservoir reduces to replacing $\varepsilon/\hbar$ with $\varepsilon/\hbar-i\Gamma/2$. For further analysis, such an approximation will be sufficient.

\section{Scattering Matrix}

The application of the above technique to scattering problems requires a number of additional steps. 

First, we need to move from imaginary Matsubara time to real time. To do this, we make an analytical continuation of expression (\ref{Gkw}) from the imaginary axis to the real one  $i\omega\to \omega+i0$ and arrive at the retarded function  $G^{(R)}(k,\omega)$. Further we use the relation between the retarded $G^{(R)}(k,\omega)$, advanced $G^{(A)}(k,\omega)$ and the casual function $G(k,\omega)$
\begin{equation} 
G(k,\omega)=\frac{1}{1-e^{-\beta\hbar\omega}}G^{(R)}(k,\omega)+\frac{1}{1-e^{\beta\hbar\omega}}G^{(A)}(k,\omega).
\end{equation} 
In the limit of low temperatures $\beta\to\infty$ $G(k,\omega)=G^{(R)}(k,\omega)$.

 Second, in the scattering theory, along with the function  $G(k,t_1-t_2)= -i\langle \mc{T}_t c_e(k,t_1)c_e^+(k,t_2)\rangle$
 we need the function $G(p,t_1;k,t_2)=\langle \mc{T}_t c_e(p,t_1)c_e^+(k,t_2)\rangle$, $\mc{T}_t$ stands for time ordering. Its Fourier component is taken on mass shell, so we can leave only the momenta as function arguments. In the Matsubara representation the function
\begin{multline}\label{Gkw1}
G(k_1,i\omega_1;k_2,i\omega_2)\\
=\left.-\left(\frac{\hbar\beta}{2\pi}\right)^2\!\!\!\!\frac{\delta^2}{\delta \bar\eta(k_1,i\omega_1)\:\delta \eta(k_2,i\omega_2)}\frac{\int \mc{D}\Theta \:e^{-\mc{A}[\eta,\bar\eta]/\hbar}}{\int \mc{D}\Theta \:e^{-\mc{A}/\hbar}}\right|_{\eta, \bar\eta=0}\!,
\end{multline}
where
\begin{multline}\label{Aeff1}
\frac{\mc{A}[\eta,\bar\eta]}{\hbar}=\frac{\mc{A}_{\rm eff}[\eta,\bar\eta]}{\hbar}
={\rm Tr}\ln {\bf M}+\frac{2\pi}{\hbar\beta}\sum_{\omega_1,\omega_2} \int  dk\\
\times \bar \eta(k_1,i\omega_1)\: G^{(0)}(k_1,i\omega_1)\delta(k_1-k_2)\delta_{\omega_1,\omega_2}\:\eta(k_2,i\omega_2),
\end{multline}
and
\begin{multline}\label{M1}
{\bf  M}=\begin{bmatrix} \left(i\omega_1-\frac{\varepsilon/2-\mu}{\hbar}\right)\delta_{0,i\omega_{12}} & \Phi(\omega_1,\omega_2)  \\ \bar\Phi(\omega_1,\omega_2) & \left(i\omega_1-\frac{-\varepsilon/2-\mu}{\hbar}\right)\delta_{0,i\omega_{12}} \end{bmatrix},\\
\Phi(\omega_1,\omega_2)=g\frac{2\pi}{\hbar\beta}\int \frac{dk}{2\pi}\:  G^{(0)}(k,i\omega_{12})\:\eta(k,i\omega_{12}),
\end{multline}
corresponds to the function  $G(k_1;k_2)$.
Here we have removed the part of the action that is not related to sources and is not needed for further analysis. At analytical continuation of imaginary frequencies on the real axis, we use the mnemonic rule $\hbar\beta\:\delta_{i\omega_1,i\omega_2}=2\pi\:\delta(\omega_1-\omega_2)$. Moreover, it should be taken into account that due to the lack of the imaginary unit in the definition of the function $G(p,t_1;k,t_2)$ an analytic continuation of the function $G(k_1,i\omega_1;k_2,i\omega_2)$ leads to the function $-iG(k_1,\omega_1+i0;k_2,\omega_2+i0)$.

In order to describe  $n$-photon scattering processes, we need the $n$-particle Green's function determined by the relation, which is a generalization of formula (\ref{Gkw1}) 
\begin{widetext}
\begin{multline}\label{Gkw2}
G(k_1,i\omega_1,\dots ,k_n,i\omega_n; k'_1,i\omega'_1,\dots, k'_n,i\omega'_n)=\\
\left.=(-1)^n\left(\frac{\hbar\beta}{2\pi}\right)^{2n}\:\frac{\delta^{2n}}{\delta \bar\eta(k_1,i\omega_1)\dots \delta \bar\eta(k_n,i\omega_n)\:\delta \eta(k'_1,i\omega'_1)\dots \delta \eta(k'_n,i\omega'_n)}\frac{\int \mc{D}\Theta \:e^{-\mc{A}[\eta,\bar\eta]/\hbar}}{\int \mc{D}\Theta \:e^{-\mc{A}/\hbar}}\right|_{\eta, \bar\eta=0},
\end{multline}
\end{widetext}
Analytical continuation of this function to the real axis for each of the Matsubara frequencies and the transition to the mass shell leads to the $n$-particle function $G^{(2n)}=G(k_1,\dots, k_n; k'_1,\dots, k'_n)$. 

Third, to calculate the connected part of the $n$-photon scattering matrix $iT^{(2n)}$, we use the LSZ reduction procedure, which consists in our case in the calculation of the  $n$-particle photon Green's function $G^{(2n)}$ and discarding external single-particle Green's functions
\begin{equation}
iT^{(2n)}=G^{(2n)} \prod_{r=1}^n[2\pi\: G_0^{-1}(k_r)G_0^{-1}(k'_r)].
\end{equation}

Let us first consider several known scattering processes in order to demonstrate how the obtained relations work.

\subsection{Single-photon scattering}
\label{SPS}
We start with the single-photon scattering matrix. Calculation by formula  (\ref{Gkw}) using (\ref{G1}) gives \cite{PhysRevB.79.205111}
\begin{multline}
G(p,k)=\left(G^{(0)}(k)\delta(k-p)-\frac{i}{2\pi}\frac{g^2}{\omega_p-\omega_\varepsilon}\left(G^{(0)}(k)\right)^2\right)\\
\times\delta(\omega_p-\omega_k),
\end{multline}
where $\omega_\varepsilon=\varepsilon/\hbar-i\Gamma/2$. Thus,
\begin{equation}\label{T1}
i\tilde T(p;k)=-i\frac{\Gamma}{v(p-k_\varepsilon)}\delta(p-k),
\end{equation}
where $k_\varepsilon=\omega_\varepsilon/v$ and symbol $\sim$ here and below denotes quantities related to  $e$-photons. Full scattering matrix $\tilde S(p;k)=\delta(p-k)+i\tilde T(p;k)$ is a quantum superposition of two processes: 1) a photon moves along a line without interacting with a TLS, 2) a photon scatters on a TLS (see Fig.~\ref{Fig_OPD}c).    

The transmission coefficient is found from the relation  $\tilde S(p;k)=\tilde t(k)\delta(p-k)$ and equal to
\begin{equation}
\tilde t(k)=\frac{k-\bar k_\varepsilon}{k-k_\varepsilon}.
\end{equation}
In the original basis $(R,L)$, the transmission and reflection coefficients of photons can be obtained using the formula for the transition between bases (\ref{RLeo}), which gives
\begin{equation}
t(k)=\frac{1}{2}(\tilde t(k)+1)=\frac{vk-\varepsilon/\hbar}{vk-\varepsilon/\hbar +i\Gamma/2},
\end{equation}
\begin{equation}
r(k)=\frac{1}{2}(\tilde t(k)-1)=\frac{-i\Gamma/2}{vk-\varepsilon/\hbar +i\Gamma/2}.
\end{equation}

\subsection{Two-photon scattering}
\begin{figure}[htp] 
	\includegraphics[width=0.1\columnwidth]{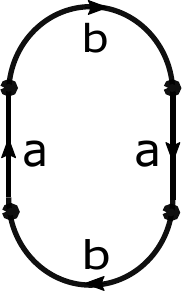}
	\caption{\small  The connected part $iT$ of the two-photon S matrix given by (\ref{T2}). 
		\label{Fig_TPD}
	}
\end{figure}

The two-photon scattering matrix can be written as
\begin{multline}
\tilde S(p_1,p_2;k_1,k_2)=\tilde S(p_1;k_1)\tilde S(p_2;k_2)+\tilde S(p_2;k_1)\tilde S(p_1;k_2)\\
+i\tilde T(p_1,p_2;k_1,k_2).
\end{multline}
To calculate $i\tilde T(p_1,p_2;k_1,k_2)$ we use expression (\ref{Aeff1}), from which we extract the forth term ($j=4$) of the ${\rm Tr}\ln {\bf M}$ expansion. Application of formula (\ref{Gkw2}) to this term leads to the diagram the TLS part of which is  shown in Fig.~\ref{Fig_TPD}.  Summing up the TLS part of the diagram over the inner Matsubara frequency, in the limit  $\beta\to \infty$ we arrive at the expression
\begin{multline}\label{G2}
G(k_1,i\omega_1,k_2,i\omega_2; k'_1,i\omega'_1,k'_2,i\omega'_2)=\frac{2g^4}{(2\pi)^4}\:\hbar\beta\\
\times\frac{(i\omega_1+i\omega_2-2\omega_\varepsilon)\:\delta_{i\omega_1+i\omega_2,i\omega'_1+i\omega'_2}}{(\omega_\varepsilon-i\omega_1)(\omega_\varepsilon-i\omega_2)(\omega_\varepsilon-i\omega'_1)(\omega_\varepsilon-i\omega'_2)}\\
\times G_0(k_1,i\omega_1)G_0(k_2,i\omega_2)G_0(k'_1,i\omega'_1)G_0(k_2,i\omega'_2).
\end{multline}

We analytically continue the expression to the real axis for each of the Matsubara frequencies, discard the external photon Green's functions, pass to the mass shell and get as a result \cite{PhysRevB.79.205111}
\begin{multline}\label{T2}
i\tilde T(p_1,p_2;k_1,k_2)\\
=i\frac{\Gamma^2}{\pi v^2}\frac{(k_1+k_2-2k_\varepsilon)\delta(k_1+k_2-p_1-p_2)}{(k_\varepsilon-k_1)(k_\varepsilon-k_2)(k_\varepsilon-p_1)(k_\varepsilon-p_2)}.
\end{multline}
Using relation (\ref{RLeo}), we arrive at an expression for the connected part of the scattering matrix for $R$-photons
\begin{multline}\label{T21}
iT(p_1,p_2;k_1,k_2)\\
=i\frac{\Gamma^2}{4\pi v^2}\frac{(k_1+k_2-2k_\varepsilon)\delta(k_1+k_2-p_1-p_2)}{(k_\varepsilon-k_1)(k_\varepsilon-k_2)(k_\varepsilon-p_1)(k_\varepsilon-p_2)}.
\end{multline}

\subsection{General case}
Similarly, one can calculate the scattering matrix for an $n$-photon process with arbitrary $n>2$, however, as $n$ increases,  expressions become cumbersome. To comprehend their general structure, we for start assume that there are only the photons of frequency $\omega_A$. Next, we recall that in the experiment 
$\omega_A-\varepsilon/\hbar=\delta\omega\ll \Gamma$, and drop $\delta\omega$ in front of $\Gamma$. We find that the $T$ matrix of the single-photon scattering  (\ref{T1}) has zero order in  $v/\Gamma$, and $T$ matrix of the two-photon scattering  (\ref{T2}) is of the second order in  $v/\Gamma$. It can be shown that in the general case of  the $(p+1)$-photon scattering $T^{(2(p+1))}\sim (v/\Gamma)^{p+1}$.  The matrix $T^{(2(p+1))}$ has dimension $k^{-(p+1)}$. The corresponding dimensionless matrix $T_{DL}^{(2(p+1))}$ is related to $T^{(2(p+1))}$ as $T_{DL}^{(2(p+1))}=(k_0)^{p+1} T^{(2(p+1))}$, where $k_0$ is the characteristic scale of momenta, for which it is reasonable to take the momentum transmitted in the scattering $\delta\omega/v$. Hence it follows that the dimensionless parameter  is the ratio $\delta\omega/\Gamma\ll 1$. For each process of $(p+1)$-photon scattering, diagrams proportional to the lowest power of this parameter will be the leading ones.
The single-photon scattering has zero order in this small parameter, so its presence as one of the channels in an $(p+1)$-photon process does not change the order of the diagram. 

Now, in addition to mode $A$ we also switch on mode $B$ and consider only those processes that go without many-particle scattering of  $B$-photons on the TLS (such processes correspond to side  peaks in the right part of the spectrum).   
\begin{widetext}
\begin{figure}[htp] 
	\includegraphics[width=1.8\columnwidth]{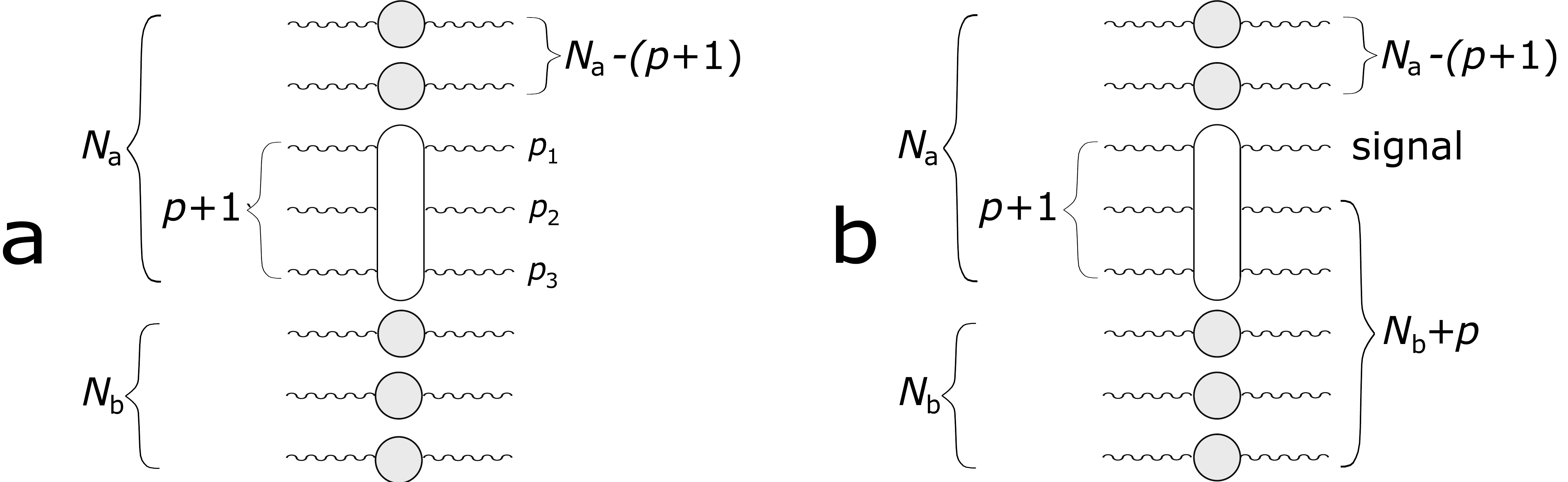}
	
	\caption{\small $(p+1)$-photon scattering from mode $A$ in the presence of photons of mode B. (a) The diagram not symmetrized with respect to scattered photons. (b) The diagram symmetrized with respect to scattered photons with momentum $p_B$.}
	\label{Fig_MD}
\end{figure}
\end{widetext}
   
  As follows from the above discussion, the leading order diagrams for  $(p+1)$-photon mode $A$ scattering have the structure shown in Fig.~\ref{Fig_MD}a. In Appendix 1 it is shown that the analytical expression for such a diagram, which we call non-symmetrized, provided that none of the scattered photons has momentum equal to $p_B$, has the form    
 \begin{equation}
\tilde S_1^{(2p+1,N_A,N_B)}=\frac{N_A!N_B!}{(p+1)!}\;i\tilde T^{(2(p+1))}\;\tilde t_A^{N_A-(p+1)}\tilde t_B^{N_B},
\end{equation}
where  $i\tilde T^{(2(p+1))}$ means the connected part of the S matrix with a corresponding set of arguments shown in Fig.~\ref{Fig_MD}a by an oval.
Of the scattered photons, a very small relative number has a momentum that lies in neighborhood of the $(p+1)$th peak. 

Another situation occurs when \textit{p} photons from $p+1$ scattered ones have momentum  $p_B$ as shown in Fig.~\ref{Fig_MD}b. We call such a diagram symmetrized. In this case, the remaining photon, indicated in Fig.~\ref{Fig_MD}b as  \textit{signal}, via the energy conservation law has a momentum corresponding to the $(p+1)$th peak. We denote the corresponding scattering matrix as $\tilde S^{(2p+1,N_A,N_B)}$ and have  
 \begin{equation}
 \tilde S^{(2p+1,N_A,N_B)}=\frac{N_A!(N_B+p)!}{(p+1)!(p+1)!}\;i\tilde T^{(2(p+1))}\;\tilde t_A^{N_A-(p+1)}\tilde t_B^{N_B}.
  \end{equation}
  It can be easily shown that in the  $(R,L)$-basis the expression for the transmitted signal goes into 
 \begin{multline}\label{St}
 S_t^{(2p+1,N_A,N_B)}\\
 =\frac{1}{2^{p+1}}\frac{N_A!(N_B+p)!}{(p+1)!(p+1)!}\;i \tilde T^{(2(p+1))}\; t_A^{N_A-(p+1)} t_B^{N_B},
 \end{multline}
  and for the reflected signal
 \begin{multline}\label{Sr}
 S_r^{(2p+1,N_A,N_B)}\\
 =\frac{1}{2^{p+1}}\frac{N_A!(N_B+p)!}{(p+1)!(p+1)!}\;i \tilde T^{(2(p+1))}\; r_A^{N_A-(p+1)} r_B^{N_B}.
 \end{multline}

\subsection{Example}
\begin{figure}[htp] 
	\includegraphics[width=0.7\columnwidth]{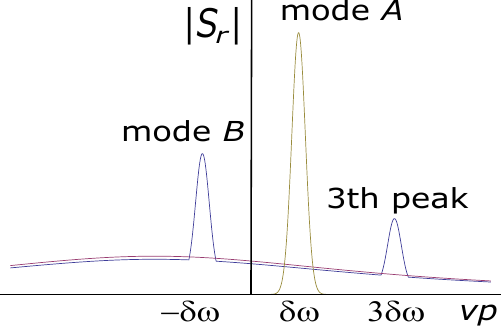}
	\caption{\small  The scattering of two photons of mode $A$ in the absence (red line) and in the presence of two photons of mode $B$ (blue line). An initial distribution of $A$-photons is depicted by the green line. The source frequency instability $\Delta\Omega$  is taken unrealistically large for the sake of clarity.
		\label{Fig_S}
	}
\end{figure}

As an example, consider the scattering of two photons of mode $A$ in the presence of two photons of mode $B$. In a unified way, all the cases discussed above are described  by the expression
 \begin{multline}
 S_r^{(2,2,2)}(p)=\frac{1}{4}\left(\frac{2!2!}{2!}\:iT(p,2k_A-p;k_A,k_A)\right.\\
 \times(1-\delta(p-k_B)-\delta(p-2k_A+k_B))+\\
 +\frac{2!3!}{2!2!}\:iT(k_B,2k_A-k_B;k_A,k_A)\\
 \left.\times(\delta(p-k_B)+\delta(p-2k_A+k_B))\right)r_B^2.
 \end{multline}
In a real experiment, the wave vector will be smeared over $\Delta k=\Delta\Omega/v$, where $\Delta\Omega$ is the scale of the radiation source frequency instability, therefore, in the above expression  $\delta$-functions will be replaced by narrow peaks of a finite width. As can be seen from Fig.~\ref{Fig_S}, the presence of only two photons in mode $B$ seriously changes the momentum distribution of scattered photons. As the occupation number $N_B$ increases, the effect of boson-stimulated scattering intensifies:   
 \begin{equation}
 \frac{ \tilde S^{(2p+1,n_A,n_B)}}{\tilde S_1^{(2p+1,n_A,n_B)}}=\frac{(N_B+p)!}{N_B!(p+1)!}\sim N_B^p\quad \text{for} 
 \quad p\sim 1,\: N_B\gg 1,
  \end{equation}
and very quickly almost all photons from mode $A$ begin to scatter into mode  $B$, which corresponds to the classical case considered in Chapter~\ref{ST}.  

\section{Side peaks}

\subsection{Amplitudes}
Now we can return to the experiment and apply the constructed formalism to the calculation of the amplitudes of the side peaks.

We can consider the modes $A$ and $B$ as two coherent states 
\begin{multline}
|\alpha\rangle=e^{-\frac{|\alpha|^2}{2}}\sum_{N_A}\frac{\alpha^{N_A}}{\sqrt{N_A!}}|N_A\rangle \\
\text{and}\quad |\beta\rangle=e^{-\frac{|\beta|^2}{2}}\sum_{N_B}\frac{\beta^{N_B}}{\sqrt{N_B!}}|N_B\rangle.
\end{multline}
Since the incoming states $|\alpha\rangle$ and $|\beta\rangle$ are uncorrelated, the general state is given by a sum of products  
\begin{equation}
|{\rm in}\rangle=e^{-\frac{|\alpha|^2+|\beta|^2}{2}}\sum_{N_A,N_B}\frac{\alpha^{N_A}\beta^{N_B}}{\sqrt{N_A!}\sqrt{N_B!}}|N_A,N_B,0 \rangle,
\end{equation}
where the third argument in the ket is the number of photons in the signal mode. As a result of he scattering on the TLS, the state $S^{(2p+1)}|{\rm in}\rangle$ arises. We are interesting in its projection on the state $|{\rm out}\rangle=|N_A-(p+1),N_B+p,1\rangle$. As a result, the amplitude of the $(2p+1)$th side peak is determined by the square of the absolute value of the expression
\begin{multline}\label{M0}
M^{(2p+1)}=e^{-\frac{|\alpha|^2+|\beta|^2}{2}}\sum_{N_A,N_B}\frac{\alpha^{N_A}\beta^{N_B}}{\sqrt{N_A!}\sqrt{N_B!}}\\
\times\langle N_A-(p+1),N_B+p,1|S^{(2p+1,N_A,N_B)}|N_A,N_B,0 \rangle.
\end{multline}
Further, writing the Fock states in the form $|N_A\rangle=(c_e^+(k_A))^{N_A}|0\rangle/\sqrt{N_A!}$ and similarly for $|N_B\rangle$ and  $|N_S\rangle$, we obtain
\begin{widetext}
\begin{equation}\label{Mp}
M^{(2p+1)}
=e^{-\frac{|\alpha|^2+|\beta|^2}{2}}\sum_{N_A,N_B}\frac{\alpha^{N_A}\beta^{N_B}}{\sqrt{N_A!}\sqrt{N_B!}}
\langle 0|\frac{a^{N_A-(p+1)}\:b^{N_B+p}\:s\: }{\sqrt{(N_A-(p+1))!}\sqrt{(N_B+p)!}}|S^{(2p+1,N_A,N_B)}|\frac{(a^+)^{N_A}\:(b^+)^{N_B}}{\sqrt{N_A!}\sqrt{N_B!}}|0\rangle.
\end{equation}
Substituting $S_r^{(2p+1,N_A,N_B)}$ from (\ref{Sr}), we arrive at the final expression
\begin{equation}\label{M2}
M_r^{(2p+1)}
=\frac{1}{2^{p+1}}\frac{i\tilde T^{(2(p+1))}}{[(p+1)!]^2}e^{-\frac{|\alpha|^2+|\beta|^2}{2}}
\sum_{N_A,N_B}\frac{\alpha^{N_A}\beta^{N_B}}{\sqrt{N_A!}\sqrt{N_B!}}\sqrt{\frac{N_A!(N_B+p)!}{(N_A-(p+1))!N_B!}}\:r_A^{N_A-(p+1)}r_B^{N_B}.
\end{equation}
\end{widetext}
If the transmitted signal is measured, then the reflection coefficients $r_A$ and $r_B$ need to be replaced by the transmission coefficients $t_A$ and $t_B$. 

\subsection{Transition to the classical limit}
Let us now proceed to calculation of side peak amplitudes in the limit of large pumps. We assume $r_A=r_B=1$. Let us recall that for the Poisson distribution, the probability of $N$ photons appearing is
\begin{equation}
P_N=|\langle N | \alpha\rangle|^2=\frac{|\alpha|^{2N}}{N!}e^{-|\alpha|^2}. 
\end{equation}
In the limit of large $N$, using the Stirling formula $N!=\sqrt{2\pi N}(N/e)^N$ and the expansion of the logarithm given in Section~\ref{Action}, it is easy to show that the probability turns into the Gaussian distribution with $|\alpha|^2=\langle N\rangle$
\begin{equation}
P_N=\frac{1}{\sqrt{2\pi\langle N\rangle}}e^{-\frac{(N-\langle N\rangle)^2}{2\langle N\rangle}},
\end{equation}
which turns into a delta function with further growth of $\langle N\rangle$. 

 Thus, from expression (\ref{M2}) we find that the amplitude of the $(2p+1)$th side peak $J_r^{(2p+1)}=|M_r^{(2p+1)}|^2$ is 
\begin{equation}\label{M3}
J_r^{(2p+1)}
=\frac{1}{2^{2(p+1)}}\frac{[\tilde T^{(2(p+1))}]^2}{[(p+1)!]^4}N_A^{p+1}N_B^p\quad\text{(r)},
\end{equation}
where we now assume $N_{A,B}=\langle N_{A,B}\rangle$.
 Substituting (\ref{T10}), we obtain
\begin{equation}\label{M4}
J_r^{(2p+1)}
=\frac{[\gamma^{(2p+1)}]^2}{\pi^p(p+1)^2}\left(\frac{v}{\Gamma}\right)^{2p}N_A^{p+1}N_B^p\quad\text{(r)}.
\end{equation}

Dimensional considerations lead to the  relation 
\begin{equation}
J_r^{(2p+1)}=N_r^{(2p+1)}v/(L^{(2p+1)}\Gamma),
\end{equation} 
where $L^{(2p+1)}$ is some characteristic length over which $(p+1)$-particle scattering occurs. The relation (\ref{M4}) transforms into $N_r^{(2p+1)}=N_A^{p+1}N_B^p$ (see (\ref{N1})) if we set 
\begin{equation}
L^{(2p+1)}=\left(\frac{\gamma^{(2p+1)}}{p+1}\right)^{1/p}L^{(1)},
\end{equation}
where $L^{(1)}=v/(\pi\Gamma)$. For several most probable processes, we obtain:  $L^{(3)}=L^{(1)}$, $L^{(5)}=\sqrt{2}L^{(1)}$, $L^{(7)}=5^{1/3}L^{(1)}$. We see that the effective interaction length gradually increases with a number of participating photons.

\section{Conclusion}
To summarize, in this paper, we undertake a consistent quantum consideration of the scattering of  radiation on a two-level system using the formalism of the scattering matrix. We  apply the developed approach to the results of the experiment on the scattering of a bichromatic drive on a two-level artificial atom. The scattering matrix in the problem under consideration has a fairly simple structure, which allowed us to establish the diagrams  that give the leading contribution to the amplitude of side  peaks observed in the experiment. We show that the spectrum observed in the experiment is the result of bosonic stimulated scattering of photons from one mode of the bichromatic drive to another and vice versa.

The developed theory can be applied as a theoretical basis for using artificial atoms as a platform for modeling many-particle phenomena.

\section{Acknowledgments}
W. V. P. acknowledges support from the RSF grant No. 23-72-30004 (https://rscf.ru/project/23-72-30004/). A. Yu. D. acknowledges the RSF grant No. 23-72-01052.

\begin{widetext}
\section{Appendix. Calculation of the multiphoton scattering diagram}
To obtain the $(N_A+N_B)$-photon Green's function, we need to calculate the following expression
\begin{equation}\label{dZ}
\frac{\delta^{\:2(N_B+N_A)}\mc{Z}}{\delta\bar\eta_S\:\underbrace{\delta\bar\eta_B\dots\delta\bar\eta_B}_{N_B+p}\:\underbrace{\delta\eta_B\dots\delta\eta_B}_{N_B}\:\underbrace{\delta\bar\eta_A\dots\delta\bar\eta_A}_{N_A-(p+1)}\:\underbrace{\delta\eta_A\dots\delta\eta_A}_{N_A}},
\end{equation}
where
\begin{multline}
\mc{Z}=e^\mc{A},\quad \mc{A}=\bar\eta_A G_A\eta_A+\bar\eta_B G_B\eta_B+G_a^{(0)}\Phi G_b^{(0)}\bar\Phi+\frac{1}{p+1}\left(G_a^{(0)}\Phi G_b^{(0)}\bar\Phi\right)^{p+1},\quad \Phi=G_A\eta_A+G_B\eta_B+G_S\eta_S.
\end{multline}
In order not to complicate the form of expressions, we have omitted all coefficients, not essential for understanding the computational process. Here the notation $G_{A,B,S}=G^{(0)}(k_{A,B,S})$ is introduced.

The first two terms in $\mc{A}$ describe the motion of photons without interaction with the TLS, the third term generates single-photon scattering processes on the TLS. Both are depicted in Fig.~\ref{Fig_MD}b by circles with adjoining wavy lines. The last term generates the connected part of the  $(p+1)$-photon process shown in the Fig.~\ref{Fig_MD}b by an oval with adjoining wavy lines. To calculate the diagram it is sufficient to take these terms in the form
\begin{equation}
G_a^{(0)}\Phi G_b^{(0)}\bar\Phi\to G_{qA}G_A^2\eta_A\bar\eta_A+G_{qB}G_B^2\eta_B\bar\eta_B,
\end{equation}
\begin{equation}
\frac{1}{p+1}\left(G_a^{(0)}\Phi G_b^{(0)}\bar\Phi\right)^{p+1}\to \frac{1}{p+1}D\: G_A^{p+1}\eta_A^{p+1}G_B^p\bar\eta_B^p\:G_S\bar\eta_S. 
\end{equation}
Here $D$ denotes the analytical expression for the TLS part of the diagram depicted by an oval, without combinatorial prefactors, the notation $G_{qA,qB}=G_q(\omega_{A,B})$ is introduced.  

Next, we need the Leibniz rule for the differentiating the product of functions 
\begin{equation}
(uv)^{(n)}=\sum_{k=0}^nC_n^k\:u^{(k)}v^{(n-k)},\quad C_n^k=\frac{n!}{(n-k)!\:k!}.
\end{equation}
We proceed to the calculation of expression (\ref{dZ}).

1. Differentiating $\mc{Z}$ once with respect to  $\bar\eta_S$, we obtain
\begin{equation}\label{W0}
\frac{1}{p+1}D\: G_A^{p+1}\eta_A^{p+1}G_B^p\bar\eta_B^p\:G_Se^\mc{A}. 
\end{equation}
Further, we can omit the last term in $\mc{A}$, since it contains a source $\bar\eta_S$, which will nullify the expressions obtained with this term when it vanishes at the end of the calculations, i. e. use  $e^{\mc{A}_1}$ instead of $e^\mc{A}$, where
\begin{equation}
 \mc{A}_1=(G_A+G_{qA}\:G_A^2)\eta_A\bar\eta_A+(G_B+G_{qB}\:G_B^2)\eta_B\bar\eta_B.
\end{equation}
2. We differentiate expression (\ref{W0}) using the Leibniz rule $(N_B+p)$ times with respect to $\bar\eta_B$, which gives
\begin{equation}
\frac{1}{p+1}D\: G_A^{p+1}\eta_A^{p+1}G_B^p G_S\sum_{k=0}^{N_B+p}C_{N_B+p}^k\frac{p!}{(p-k)!}\bar\eta_B^{p-k}\left[(G_B+G_{qB}\:G_B^2)\eta_B\right]^{N_B+p-k}e^{\mc{A}_1}. 
\end{equation}
Here and below we denote  $G_{qA}=G_q(\omega_A)$, $G_{qB}=G_q(\omega_B)$. The only term that does not vanish when  $\bar\eta_B$ vanishes is a term with $k=p$ equal to 
\begin{equation}\label{W1}
\frac{p!}{p+1}C_{N_B+p}^p\:D\: G_A^{p+1}\eta_A^{p+1}G_B^p\:G_S\:(G_B+G_{qB}\:G_B^2)^{N_B}\eta_B^{N_B}\:e^{\mc{A}_1}.
\end{equation}

3. Differentiating (\ref{W1})  $N_B$ times with respect to $\eta_B$, we obtain 
\begin{equation}\label{W2}
\frac{p!N_B!}{p+1}C_{N_B+p}^p\:D\: G_A^{p+1}\eta_A^{p+1}G_B^p\:G_S\:(G_B+G_{qB}\:G_B^2)^{N_B}\:e^{\mc{A}_1}.
\end{equation}

4. We differentiate (\ref{W2}) $(N_A-(p+1))$ times with respect to $\bar\eta_A$, which leads to 
\begin{equation}\label{W3} 
\frac{p!N_B!}{p+1}C_{N_B+p}^p\:D\: G_A^{p+1}\eta_A^{N_A}G_B^p\:G_S\:(G_B+G_{qB}\:G_B^2)^{N_B}(G_A+G_{qA}\:G_A^2)^{N_A-(p+1)}\:e^{\mc{A}_1}.
\end{equation}

5. We differentiate  (\ref{W3}) $N_A$ times with respect to  $\eta_A$, using the Leibniz rule and obtain
\begin{multline}
\frac{p!N_B!}{p+1}C_{N_B+p}^p\:D\: G_A^{p+1}G_B^p\:G_S\:(G_B+G_{qB}\:G_B^2)^{N_B}(G_A+G_{qA}\:G_A^2)^{N_A-(p+1)}\times\\
\times\sum_{k=0}^{N_A}C_{N_A}^k\frac{N_A!}{(N_A-k)!}\eta_A^{N_A-k}(G_A+G_{qA}\:G_A^2)^{N_A-k}\:e^{\mc{A}_1}.
\end{multline}
The only term that does not vanish when $\eta_A$ and $\bar\eta_A$ vanish is the term with $k=N_A$ equal to 
\begin{equation}
\frac{p!N_B!N_A!}{p+1}C_{N_B+p}^p\:D\: G_A^{p+1}G_B^p\:G_S(G_B+G_{qB}\:G_B^2)^{N_B}(G_A+G_{qA}\:G_A^2)^{N_A-(p+1)}\:e^{\mc{A}_1}.
\end{equation}
Substituting $C_{N_B+p}^p$ and setting equal to zero all sources, we arrive at the final expression
\begin{equation}
\frac{N_A!(N_B+p)!}{p+1}\:G_A^{p+1}D\:G_S G_B^p\:(G_B+G_{qB}\:G_B^2)^{N_B}(G_A+G_{qA}\:G_A^2)^{N_A-(p+1)}.
\end{equation}

Now, to pass to the scattering matrix, we perform the LSZ reduction procedure. By omitting the functions $G_A^{p+1}$, $G_S$, and $G_B^p$, we remove the lines adjacent to the oval. To each of the last two expressions in parentheses we assign an expression of the form $1+iT$, which in turn is equal to the transmission coefficient $\tilde t$. As a result, we get
\begin{equation}\label{S10}
\frac{N_A!(N_B+p)!}{p+1}D\:\tilde t_B^{N_B}\:\tilde t_A^{N_A-(p+1)}.
\end{equation}

Now let us calculate the numerical prefactor of the diagram, and also refine the form of $D$. We start with a connected diagram, for which $N_A=p+1$, $N_B=0$.
 
 In the zeroth approximation with respect to parameter $\delta\omega/\Gamma$ for $D=D^{(2p+1)}=G_{qA}^{(p+1)}G_{qB}^{(p+1)}$  the following expression is valid:  
\begin{equation}
D^{(2p+1)}=-i\:2^{2p+1}\:2\pi\:\gamma^{(2p+1)}\left(\frac{v}{\Gamma}\right)^p.
\end{equation}
The factor $2\pi$ in this expression arises after a single integration over the frequency running around the loop. The numerical factor $\gamma^{(2p+1)}$ arises when calculating a specific diagram. Unfortunately, we do not know  a general formula, however, it can be easily found for each specific diagram, here are the first few values: $\gamma^1=1$, $\gamma^3=2$, $\gamma^5=-6$, $\gamma^7=-20$. 

The whole diagram also contains the factor $(2\pi)^{-2(p+1)}$ arising from the matrix $(G_a^{(0)}\Phi G_b^{(0)}\bar\Phi)^{p+1}$. The factor  $(2\pi)^{(p+1)}$ arises as a result of the LSZ reduction procedure. 

When calculating $G_{2(p+1)}$ by formula (\ref{Gkw2}), the factor $(-1)^{p+1}$ arises, and it is necessary to take into account the minus sign of the action in $e^{-\mc{A}/\hbar}$, i. e. the resulting sign is  $(-1)^p$.

The analytic continuation gives rise to the factor $i(-i)^{2(p+1)}=(-1)^p i$.

For the sake of completeness, we list here again the factors already present in expression (\ref{S10}). This is the coefficient arising from the expansion of ${\rm Tr}\ln$ equal to $2/(2(p+1))$. The factor two in the numerator is the result of taking the trace, the denominator arises from the expansion of the logarithm into a series. In addition, the connected diagram contains a combinatorial factor  $(p+1)!(p+1)!$ corresponding to the numerator of the fraction in  (\ref{S10}). Collecting all the factors, we arrive at the expression for the connected diagram
\begin{equation}\label{T10}
i\tilde T^{(2(p+1))}\approx -\frac{1}{(2\pi)^p}\frac{(p+1)!(p+1)!}{p+1}\:2^{2p+1}\gamma^{(2p+1)}\left(\frac{v}{\Gamma}\right)^p.
\end{equation}
We recall that the factor $2^{-(p+1)}$ arises when passing between bases.

The general case of a disconnected diagram with  $N_A\ge p+1$, $N_B\ge 0$ differs from the one discussed above only because the factor $(p+1)!(p+1)!$ is replaced by $N_A!(N_B+p)!$ in the case of symmetrization with respect to $(N_B+p)$ output photons, and by $N_A!(p+1)!N_B!$ in the absence of symmetrization. Thus, 
\begin{equation}
 \tilde S^{(2p+1,N_A,N_B)}\approx -\frac{1}{(2\pi)^p}\:2^{2p+1}\gamma^{(2p+1)}\left(\frac{v}{\Gamma}\right)^p\tilde t_B^{N_B}\:\tilde t_A^{N_A-(p+1)}\times\left\{\begin{aligned}\frac{N_A!\:(N_B+p)!}{p+1}&\quad\text{ symmetrized diagram,}\\\frac{N_A!\:(p+1)!\:N_B!}{p+1}&\quad\text{non-symmetrized diagram.}\end{aligned}\right.
\end{equation}
\end{widetext}

\bibliography{sample}

%

\end{document}